# Conformational Dynamics of Supramolecular Protein Assemblies in the EMDB

Do-Nyun Kim, Cong-Tri Nguyen, Mark Bathe[*]
Department of Biological Engineering
Massachusetts Institute of Technology

**ABSTRACT**

The Electron Microscopy Data Bank (EMDB) is a rapidly growing repository for the dissemination of structural data from single-particle reconstructions of supramolecular protein assemblies including motors, chaperones, cytoskeletal assemblies, and viral capsids. While the static structure of these assemblies provides essential insight into their biological function, their conformational dynamics and mechanics provide additional important information regarding the mechanism of their biological function. Here, we present an unsupervised computational framework to analyze and store for public access the conformational dynamics of supramolecular protein assemblies deposited in the EMDB. Conformational dynamics are analyzed using normal mode analysis in the finite element framework, which is used to compute equilibrium thermal fluctuations, cross-correlations in molecular motions, and strain energy distributions for 452 of the 681 entries stored in the EMDB at present. Results for the viral capsid of hepatitis B, ribosome-bound termination factor RF2, and GroEL are presented in detail and validated with all-atom based models. The conformational dynamics of protein assemblies in the EMDB may be useful in the interpretation of their biological function, as well as in the classification and refinement of EM-based structures.

[*]Corresponding author:
Mark Bathe
77 Massachusetts Avenue
Building NE47, Room 223
Cambridge, MA 02139, U.S.A.
Tel. +1–617–324–5685
Fax +1–617–324–7554
mark.bathe@mit.edu

**INTRODUCTION**

Single-particle reconstructions of supramolecular protein assemblies deposited in the publically accessible Electron Microscopy Data Bank (EMDB http://www.emdatabank.org/ and http://www.ebi.ac.uk/pdbe/emdb/) have been growing rapidly in recent years, representing a total of approximately 250 distinct structures in 2009 [1, 2]. Recent growth of the EMDB parallels early growth of the Protein Data Bank (PDB), which has developed to include tens of thousands of protein crystal structures since its inception in 1971 [3, 4]. While the static structure of proteins provides invaluable insight into their biological function, their conformational dynamics often play an additional important role in understanding their function mechanistically.

Normal mode analysis (NMA) has proven to be an effective computational approach to investigate biologically relevant collective motions about a representative ground-state structure, or ensemble thereof [5]. The primary advantage of NMA over molecular dynamics is its relative computational efficiency, which is a result of the harmonic approximation of atomic motions about the ground-state conformation, as well as the neglect of explicit solvent degrees of freedom. Computational efficiency is further enhanced in NMA by using coarse-grained modeling approaches that reduce the number of protein degrees of freedom, which has been essential to facilitating the analysis of high molecular weight supramolecular assemblies. Popular approaches include the Rotational Translational Blocks (RTB) procedure [6], which requires atomic coordinates for the underlying protein structure, the Elastic Network Model [7], and more recently the Finite Element Method (FEM) [8]. The FEM provides a natural framework for the computation of conformational dynamics and mechanics of high molecular weight proteins and their assemblies based on EM reconstructions because the FE model is defined using its closed molecular surface, which is naturally provided by single-particle reconstructions. In the FE framework, proteins are modeled as homogeneous isotropic elastic materials characterized by a mean mass density and elastic stiffness.

While several data banks and servers [9-14] exist to disseminate publically the conformational dynamics of protein structures deposited in the PDB, similar data banks do not exist at present for the EMDB. Such data bank would support both further computational analyses to gain mechanistic insight into the biological function of these assemblies, as well as potentially serve as a basis set for classification in single-particle reconstruction. Toward this end, here we present the Electron Microscopy Normal Modes Data Bank (EM-NMDB) to provide conformational dynamics of structures present in the EMDB. The FE framework is used to calculate the lowest normal modes of all structures for which a well defined molecular surface can be calculated based either on the EM density contour level or molecular weight provided in the entry [8]. The EM-NMDB includes normal mode shapes and their root-mean-square amplitudes in thermal equilibrium, their equilibrium mechanical strain energy, and the discretized molecular surfaces used to perform the analyses. Results for the viral capsid of hepatitis B, ribosome-bound termination factor RF2, and GroEL are presented in detail, including quantitative comparison of the lowest normal modes with those of atomic-level models, equilibrium thermal fluctuations, and correlations in dynamical motions between distant domains. Effects of EM resolution are additionally examined. Results for the EM-NMDB are available at http://lcbb.mit.edu/~em-nmdb/.

**METHODS**

The EM-NMDB is maintained through an automated procedure that consists of several distinct computational steps (Figure 1): (1) retrieval of the EM density map; (2) molecular surface computation and discretization; (3) discretized molecular surface evaluation and repair; (4) finite element mesh generation and normal mode analysis; and (5) results processing. The EM-NMDB server monitors the EMDB regularly to determine when new structures suitable for conformational dynamics analysis have been deposited. To date, 594 maps out of 681 EMDB entries have been downloaded and analyzed successfully. Of the remaining depositions, 55 maps are on hold by the EMDB, one map file is broken (unreadable), and 31 maps are tomograms.

For computation of the molecular surface in step 2, the suggested contour level provided by the EMDB is used unless no such contour level is provided. In this case, the molecular weight is used instead (four entries), where the contour level corresponding to the given molecular volume assuming a protein mass density of 1.35 g/cm$^3$ is employed [15]. If neither the contour level nor the molecular weight of the complex is provided, the EMDB entry is classified as "molecular surface indeterminable" and no analysis is performed (ten entries, **Table S1**). Additionally, in several cases structures consisting of disconnected multiple bodies are obtained using the suggested contour level, in which case the entry is flagged as having "disconnected multiple bodies" and no analysis is performed (87 entries, Table S1). Several examples of such maps are presented in Supplementary Material (Figure S1 and Table S1). Discretization of the molecular surface is performed using the marching cubes algorithm [16] implemented in Chimera [17]. The triangulated surface is subsequently exported in OBJ format, a geometry definition file format originally developed by Wavefront Technologies, Inc., Santa Barbara, CA.

In cases where both the suggested contour level and the molecular weight are provided, the molecular volume enclosed by the molecular surface discretized by the FE mesh ($V_{FE}$) is compared with the molecular volume predicted by the molecular weight provided ($V_{MW}$) assuming a typical protein mass density of 1.35 g/cm$^3$. The relative difference between these values, $\Delta V_{rel} = \left|1 - {V_{MW}}/{V_{FE}}\right| \times 100$ (%), is computed and denoted using 10 for $0 \leq \Delta V_{rel} < 10\%$, 20 for $10\% < \Delta V_{rel} < 20\%$, 50 for $20\% < \Delta V_{rel} < 50\%$, 100 for $50\% \leq \Delta V_{rel} < 100\%$ and 100+ for $\Delta V_{rel} > 100\%$.

In general, the triangulated molecular surface generated using Chimera is not closed and contains small isolated fragments where an "isolated fragment" is defined to be a closed surface that consists of fewer than 10% of the number of triangular faces that forms the largest structure in the map. It additionally often contains intersecting, overlapping, degenerate, and/or non-manifold surface triangles. Because the FEM requires unique, closed surfaces for the generation of a volumetric mesh, surface mesh repair is required in step 3 prior to performing FE-based NMA (Figure 2). Surface mesh filters available in Meshlab [18] are used for this purpose. Meshlab reads the OBJ file format exported from Chimera and exports the filtered molecular surface in STL file format, which is native to the stereolithography CAD software created by 3D Systems, Inc., Rock Hill, SC. The resulting STL file from Meshlab is imported to the commercially available FEA program ADINA (ADINA R&D, Inc., Watertown, MA), which is used to generate the 3D FE volume mesh consisting of 4-node tetrahedral finite elements [19]. If mesh repair is impossible using Meshlab, then the EMDB entry is classified as "failed in molecular surface repair" and no further analysis is performed (45 entries, **Table S1**).

The mesh filtering and repair scheme employs several filters available in Meshlab. The original surface mesh obtained from Chimera is first processed using basic filters with default parameters that remove duplicate faces, unreferenced vertices, zero-area faces, self-intersecting faces, isolated fragments, and non-manifold faces. Default parameters are additionally used to close holes that are in the original surface mesh and are created by removing defective faces. In case where closing holes re-introduce problems with the surface, the surface mesh is successively refined, smoothened and coarsened where "Midpoint subdivision [20]", "Laplacian smooth [21]" and "Quadratic edge collapse decimation [22]" are used for refinement, smoothing, and coarsening, respectively.

In step 3, FE analysis is used to calculate the lowest twenty normal modes based on the three-dimensional volume mesh. While twenty normal modes are chosen for the initial database because they generally describe approximately 80% of the total magnitude of equilibrium thermal fluctuations (Figure S3), additional normal modes may easily be calculated using the FE model as required. Proteins are modeled as homogeneous linear isotropic materials characterized by three independent effective material parameters: the Young's modulus ($E$), the mass density ($\rho$), and Poisson's ratio ($\nu$), where proteins are assumed to have mass density 1.35 g/cm$^3$ and Poisson's ratio 0.3 [23], which is typical of crystalline solids. While the effective Young's modulus is generally unknown for proteins, it can be obtained by fitting thermal fluctuations of α-carbon atoms in the FE model to those obtained using either the all-atom normal mode analysis or the RTB procedure when atomic coordinates are available, which generally ranges from 2 GPa to 5 GPa [8, 24]. Because most structures in the EMDB lack atomic coordinates, normal mode amplitudes and dependent properties are computed using a Young's modulus of 2 GPa, representing a lower bound on the protein stiffness [8]. The precise value of the Young's modulus affects linearly the magnitude of thermal fluctuations, and therefore all results presented may be scaled linearly to calculate their value corresponding to higher or lower Young's moduli.

The subspace iteration procedure [24, 25] is used to solve the eigenvalue problem using $2N_m$ starting iteration vectors, where $N_m$ denotes the number of eigenmodes to be calculated.

The computed number of rigid body modes is compared with the number of isolated molecular volumes calculated in the surface discretization step by successively removing the largest components from the molecular surface until no component remains and counting the number of those steps, where each isolated fragment has six rigid body modes.

Each normal mode amplitude is scaled according to the equipartition theorem [26], which requires that the equilibrium mean elastic energy associated with each normal mode equals $\frac{1}{2}k_B T$, where $k_B$ is the Boltzmann constant and $T$ is temperature, taken here to be 300 K. The equilibrium mean elastic energy associated with each mode $k$ is given by $\langle \mathcal{V}_k \rangle = \langle \frac{1}{2}(\alpha_k \mathbf{x}_k)^T \mathbf{K}(\alpha_k \mathbf{x}_k) \rangle = \frac{1}{2}\langle \alpha_k^2 \lambda_k \rangle = \frac{1}{2}k_B T$ where $\mathbf{x}_k$ denotes the mass normalized eigenvector, $\alpha_k$ is its amplitude, $\mathbf{K}$ is the stiffness matrix and $\lambda_k$ is the eigenvalue associated with mode $k$. Scaled normal modes are interpolated to each voxel of the original density map and stored in ASCII format. Eigenvector magnitudes are stored in the MRC density map format so that both the original density map and the eigenvector magnitude map may be viewed at the same time (for example using Chimera, Figure S2). The root mean squared fluctuation (RMSF) amplitude are computed at each FE nodal point using the lowest twenty normal modes (Figure S3), and are interpolated to each voxel to be stored in ASCII format. The mean squared fluctuation for FE node $i$ is given by $\langle \Delta r_i^2 \rangle = \sum_k \langle \Delta r_{ik}^2 \rangle = \sum_k \alpha_k^2 x_{ik}^2 / m_i$ where $m_i$ denotes the effective mass of node $i$, which is assumed to be same for all FE nodal points. In addition, initial and deformed molecular surfaces for each mode are stored in STL file format for use in programs such as Maya (Autodesk, Inc., San Rafael, CA). Molecular animations in high (640x320 pixels) and low (320x160 pixels) resolutions are provided for four sub-frames in orthogonal views: ISO-3D, XY-plane, XZ-plane and YZ-plane to illustrate the dynamical motions associated with these modes.

*Normal mode validation*

EM-based normal modes computed using the FEM are compared with all-atom based results using the molecular surface from the atomic crystal structure for several validation cases, including the viral capsid of hepatitis B and GroEL. Similarity in normal modes is computed using the symmetric mean overlap, $O_i^{a,b} = 0.5\left(\max_{i-ne \leq j \leq i+ne} P_{ij}^{a,b} + \max_{i-ne \leq j \leq i+ne} P_{ij}^{b,a}\right)$ for mode $i$ where $P_{ij}^{a,b} = \left|x_i^a \cdot x_j^b\right| / \left|x_i^a\right|\left|x_j^b\right|$, $ne$ is the number of neighboring modes used to account for potential mode-swapping between models with closely spaced eigenvalues, $a$ and $b$ denote the different models to be compared, and $x_i^a$ represents the eigenvector of the $i$-th mode [27].

*Correlations in molecular motions*

The Linearized Mutual Information (LMI) is used to calculate correlations in molecular motions [28]. The Mutual Information (MI) in atomic displacements is defined as $\mathcal{I}\left[\Delta r_1, \Delta r_2, \cdots, \Delta r_N\right] = \int p(\Delta r) \ln \frac{p(\Delta r)}{\prod_{i=1}^{N} p_i(\Delta r_i)} d\Delta r$ where $\Delta r_i$ denotes the positional fluctuation vector of atom $i$, and $p_i(\Delta r_i)$ and $p(\Delta r)$ are their marginal and joint probability distributions, respectively. The LMI can be written in terms of equilibrium conformational properties as $\mathcal{I}_{lin}\left[\Delta r_i, \Delta r_j\right] = \frac{1}{2}\left(\ln \det \mathbf{C}_{(i)} + \ln \det \mathbf{C}_{(j)} - \ln \det \mathbf{C}_{(ij)}\right)$ where $\mathbf{C}_{(ij)} = \left\langle \left(\Delta r_i, \Delta r_j\right)^T \left(\Delta r_i, \Delta r_j\right)\right\rangle$ and $\mathbf{C}_{(i)} = \left\langle \Delta r_i^T \Delta r_i \right\rangle$. The generalized correlation coefficient, $r_{LMI}\left[\Delta \mathbf{r}_i, \Delta \mathbf{r}_j\right] = \left\{1 - \exp\left(-2\mathcal{I}_{lin}\left[\Delta \mathbf{r}_i, \Delta \mathbf{r}_j\right]/d\right)\right\}^{1/2}$, is employed, where $d$ is dimensionality ($d = 1, 2, 3$) [28]. Accordingly, LMI components are computed in the standard way using the normal modes and $\left\langle \Delta \mathbf{r}_i^T \Delta \mathbf{r}_j \right\rangle = \frac{k_B T}{\sqrt{m_i m_j}} \sum_{k=1}^{nm} \frac{\mathbf{x}_{ik}^T \mathbf{x}_{jk}}{\lambda_k}$ where $m_i$ is the mass of atom $i$, $nm$ is the number of normal modes and $\mathbf{x}_{ik}$ is the displacement vector of atom $i$ due to mode $k$ [29, 30]. The MI metric is employed due to its higher sensitivity in detecting correlations in molecular motions than the more commonly used Pearson correlation coefficient, which does not account for non-colinear correlated motions [28, 30].

To identify molecular domains that are highly correlated in motion, hierarchical clustering is performed using $1 - r_{LMI}$ as a distance metric [31] and defining the distance between clusters as the average distance between all pairs of atoms in any two clusters. Although this process naturally forms clusters of atoms with respect to their magnitude of MI, in most cases clusters with high MI consist of atoms that are spatially near one another because direct bonded/contact interactions introduce highly correlated motions, which are trivial and not of interest. Instead, we seek to identify atomic clusters that are both highly correlated and spatially distant, which could not be identified from molecular structure/geometry alone. To achieve this, $N$ clusters are formed such that the Kelley–Gardner–Sutcliffe (KGS) penalty function reaches at its minimum [32], which measures the balance between the average variation within each cluster and across all clusters. Then each cluster is divided into its two sub-clusters and checked to determine whether they are composed of "distant" sub-clusters using the criterion, $D_{ij} \geq C_{dist} \times \left(R_{g,i} + R_{g,j}\right)$ where $D_{ij}$ is the distance between the mean positions of (sub-)clusters $i$ and $j$, $R_{g,i}$ is the radius of gyration of cluster $i$, and $C_{dist}$ is an empirical parameter that is used to define "distant." Among pairs of distant clusters, the pair with the highest mean correlation is chosen. If no distant sub-clusters exist based on the above criterion, the same procedure is applied to the original clusters to identify distant clusters that are highly correlated. $C_{dist} = 1.2$ is taken as the default value. This approach is tested for T4 lysozyme (PDB ID 3LZM) and Adenylate kinase (PDB ID 4AKE, open conformer). Residue clusters obtained for T4 lysozyme correspond to residues correlated due to hinge-bending [8, 28] and those for Adenylate kinase are active residues in the conformational change from its open (PDB ID 4AKE) to its closed states (PDB ID 1AKE) [24, 33] (Figure S4 and Figure S5).

*EM-NMDB Statistics*

The preceding analysis approach is applied to 497 EMDB entries. 452 structures are solved, where 45 structures failed in our molecular surface repair procedure (Figure 3 and **Table S1**). Proteins are classified according to biological function by title and sample name key words provided in the EMDB (Figure 4). The EMDB covers a range of proteins including viruses as the dominant class, and RNA binding proteins and protein kinases as major subclasses of protein complexes.

**RESULTS**

Conformational dynamics of structures in the EMDB are publically accessible through the EM-NMDB server (http://lcbb.mit.edu/~em-nmdb). Results include normal modes and their equilibrium magnitudes, triangulated molecular surfaces and the deformed surface meshes, the RMSF and the elastic strain energy distributions. Results for the lowest twenty modes for each structure are currently deposited, with the possibility of increasing the number in the near future.

The RMSFs are also computed and stored using the lowest 20 normal modes (Figure 5). These results may be used to provide insight into biologically relevant motions or to suggest alternate conformations that may be used in multi-reference refinement in single-particle reconstruction [34]. The elastic strain energy distributions as well as displacements for each mode are also presented, potentially identifying domain motions that correspond to high deformation energies in flexible regions linking rigid domains. Results are presented for hepatitis B virus, ribosome-bound termination factor RF2, and GroEL, where the effect of EM resolution on the normal modes and associated conformational properties are also evaluated for these structures [35, 36].

*Hepatitis B virus*

Hepatitis B virus (HBV) is a small virus that consists of a nucleocapsid core enclosing DNA and an outer lipid envelope, and is physiologically relevant to liver infection in humans [37]. The HBV capsid has icosahedral symmetry consisting of 240 subunits in a T = 4 state, where T denotes the triangulation number [38], with 12 pentameric and 30 hexameric capsomers. Cryo-EM-based normal mode results for the HBV nucleocapsid (EMD-1402) are compared with the reference model obtained from the atomic structure, where the molecular surface is given by the global density field that is a superposition of the densities of each atom constructed using a Gaussian density distribution [39]. The EM-based FE model consists of 18,766 nodes and 72,898 tetrahedra while the reference FE model contains 36,985 nodes and 173,621 tetrahedra (Figure 6). The mean modal overlap (Figure 7) shows that the lowest normal modes are robust to variations in molecular surface resolution, as found previously [40]. RMSFs in molecular motions suggest that hexameric capsomers are more flexible than their pentameric counterparts (Figure 8). The relative rigidity of pentamers may be attributed to the structural fact that they lack pores.

### *Ribosome-bound termination factor RF2*

Release factors (RFs) are proteins that recognize a stop codon that terminates protein synthesis when it arrives at the ribosomal decoding center [41, 42]. The EM-based FE model for ribosome-bound termination factor RF2 (EMD-1010) is constructed with 1,509 nodal points and 6,155 tetrahedral elements (Figure 9A). A region of high RMSFs (Figure 9B) corresponds to a conserved GGQ amino-acid motif that is thought to be important for peptide release by interacting directly with the peptidyl-transferase centre (PTC) [41]. A major conformational shape change is also observed in this region between the wild type and a mutant that inhibits peptide release [41]. Seven compact clusters are obtained using the KGS criterion (Figure 10). Maximally correlated distant clusters consist of two domains that exhibit highly correlated collective motions when the closed-state crystal structure is fit into the corresponding cryo-EM map in the open state [41].

### *GroEL*

GroEL is an extensively studied bacterial chaperonin that consists of two rings composed of seven identical subunits [43-48]. Three domains form one subunit: the equatorial domain that connects the two rings and where ATP binds, the apical domain where ligands bind, and the intermediate domain connecting the equatorial and apical domains.

GroEL is analyzed at 6 Å (EMD-1081), 11.5 Å (EMD-1080), and 25 Å (EMD-1095) resolutions and results are compared with the reference atomic model based on the crystal structure (PDB ID 1OEL). The molecular surface of the reference model is obtained from MSMS using a 1.4 Å radius probe [49]. FE models and their mean mode overlaps (Figure 11,

Figure 12, and Table 1) demonstrate a lack of sensitivity of the lowest normal modes to molecular surface resolution [40]. This high degree of similarity in the lowest normal modes is attributable to the fact that the overall shape determines the global/collective motions of proteins [8, 40]. However, unlike the low resolution models, the high resolution, 6 Å EM-based model of GroEL contains bending motions of the two rings mixed with their stretching and shearing motions, which deteriorates the mean overlap slightly. This is due to the fact that the suggested contour level provided in the EMDB results in considerably weaker connection between the rings as compared with the other models, including the reference atomic model (Figure S6).

To investigate correlations in molecular motions, the lowest 20 normal modes of the 11.5 Å EM-based GroEL model are projected onto the alpha carbon atoms of the reference atomic structure and the generalized correlation coefficient computed, $r_{LMI}$. Hierarchical clustering with 28 clusters leads to the division of each subunit into two clusters, one with the equatorial domain and the other with the apical and intermediate domains (top-right of Figure 13). Among adjacent clusters (data points in the horizontally shaded area in Figure 13), those in neighboring subunits within the same ring or between two rings show the highest correlations (Figure 13A-B), while clusters within one subunit are somewhat less correlated (Figure 13C). Additionally, next-nearest clusters either in the same ring or in different rings (Figure 13D-E) are as correlated as clusters within one subunit (data points in the vertically shaded area in Figure 13). These correlated intra- and inter-ring interactions are important to the allosteric mechanism of GroEL, corresponding to positive and negative cooperativity, respectively, in the binding and hydrolysis of ATP [43, 45, 48].

**CONCLUSIONS**

An unsupervised computational framework based on the FEM is presented to compute the conformational dynamics of supramolecular assemblies of unknown atomic structures deposited in the EMDB. The FE-based approach is a natural choice for computation of NMs from molecular surface representations such as obtained from EM density maps due to the fact that it requires a closed surface to form the dynamical equations of motion governing the protein's conformational dynamics. The procedure developed here is adequate for the majority of EMDB structures that have a density level that leads to a well defined molecular surface. Results are publically accessible via a data bank that is configured to update automatically when new structures are deposited in the EMDB, and are expected to be useful in understanding biologically relevant functional motions of EM-based structures.

**ACKNOWLEDGEMENTS**

Useful discussions with Gaël McGill, Christoph Best, Janet Iwasa, Wah Chiu, Joachim Frank, Andres Leschziner, Grant Jensen, Martin Karplus, and Qiang Cui are gratefully acknowledged. Funding for this work was provided by MIT Faculty Start-up Funds and the Samuel A. Goldblith Career Development Professorship awarded to MB.

# REFERENCES

1. Tagari, M., et al., *New electron microscopy database and deposition system.* Trends In Biochemical Sciences, 2002. **27**(11): p. 589-589.
2. Henrick, K., et al., *EMDep: a web-based system for the deposition and validation of high-resolution electron microscopy macromolecular structural information.* Journal Of Structural Biology, 2003. **144**(1-2): p. 228-237.
3. Bernstein, F.C., et al., *The Protein Data Bank: A Computer-based Archival File for Macromolecular Structures.* Journal of Molecular Biology, 1977. **112**(3): p. 535-542.
4. Berman, H.M., et al., *The Protein Data Bank.* Nucleic Acids Research, 2000. **28**(1): p. 235.
5. Cui, Q. and I. Bahar, eds. *Normal Mode Analysis: Theory and Applications to Biological and Chemical Systems*. Mathematical and Computational Biology Series. 2006, Chapman & Hall/CRC: Boca Raton. 406.
6. Tama, F., et al., *Building-block approach for determining low frequency normal modes of macromolecules.* Proteins: Structure Function and Genetics, 2000. **41**(1): p. 1-7.
7. Bahar, I. and A.J. Rader, *Coarse-grained normal mode analysis in structural biology.* Current Opinion in Structural Biology, 2005. **15**(5): p. 586-592.
8. Bathe, M., *A Finite Element framework for computation of protein normal modes and mechanical response.* Proteins: Structure, Function and Bioinformatics, 2008. **70**(4): p. 1595-1609.
9. Alexandrov, V., et al., *Normal modes for predicting protein motions: A comprehensive database assessment and associated Web tool.* Protein Science, 2005. **14**(3): p. 633.
10. Suhre, K. and Y.-H. Sanejouand, *ElNemo: a normal mode web server for protein movement analysis and the generation of templates for molecular replacement.* Nucleic Acids Research, 2004. **32**(suppl_2): p. W610-614.
11. Wako, H., M. Kato, and S. Endo, *ProMode: a database of normal mode analyses on protein molecules with a full-atom model.* Bioinformatics, 2004. **20**(13): p. 2035-2043.
12. Yang, L.-W., et al., *iGNM: a database of protein functional motions based on Gaussian Network Model.* Bioinformatics, 2005. **21**(13): p. 2978-2987.
13. Hollup, S.M., G. Salensminde, and N. Reuter, *WEBnm@: a web application for normal mode analyses of proteins.* BMC Bioinformatics, 2005. **6**: p. Art. No. 52.
14. Lindahl, E., et al., *NOMAD-Ref: visualization, deformation and refinement of macromolecular structures based on all-atom normal mode analysis.* Nucleic Acids Research, 2006. **34**: p. W52-W56.
15. Fischer, H., I. Polikarpov, and A.F. Craievich, *Average protein density is a molecular-weight-dependent function.* Protein Science, 2004. **13**(10): p. 2825-2828.
16. Lorensen, W.E. and H.E. Cline, *Marching cubes: A high resolution 3D surface construction algorithm.* Computer Graphics (SIGGRAPH '87 Proceedings), 1987. **21**(4): p. 163-169.
17. Pettersen, E.F., et al., *UCSF chimera - A visualization system for exploratory research and analysis.* Journal Of Computational Chemistry, 2004. **25**(13): p. 1605-1612.
18. Cignoni, P., et al., *Meshlab: An Open-source Mesh Processing Tool*, in *Sixth Eurographics Italian Chapter Conference*. 2008. p. 129-136.
19. Béchet, E., J.C. Cuilliere, and F. Trochu, *Generation of a finite element MESH from stereolithography (STL) files.* Computer-Aided Design, 2002. **34**(1): p. 1-17.

20. Peters, J. and U. Reif, *The simplest subdivision scheme for smoothing polyhedra.* ACM Trans. Graph., 1997. **16**(4): p. 420-431.
21. Field, D.A., *Laplacian smoothing and Delaunay triangulations.* Communications in Applied Numerical Methods, 1988. **4**(6): p. 709-712.
22. Garland, M. and P.S. Heckbert. *Surface simplification using quadric error metrics*. in *SIGGRAPH*. 1997.
23. Kharakoz, D.P., *Protein compressibility, dynamics, and pressure.* Biophysical Journal, 2000. **79**(1): p. 511-525.
24. Sharifi Sedeh, R., M. Bathe, and K.J. Bathe, *The subspace iteration method in protein normal mode analysis.* Journal of Computational Chemistry, 2009. **31**(1): p. 66-74.
25. Bathe, K.J. and S. Ramaswamy, *An accelerated Subspace Iteration Method.* Computer Methods in Applied Mechanics and Engineering, 1980. **23**(3): p. 313-331.
26. McQuarrie, D.A., *Statistical Mechanics.* Harper's Chemistry Series. 1975, New York: Harper & Row.
27. Kondrashov, D.A., et al., *Protein structural variation in computational models and crystallographic data.* Structure, 2007. **15**(2): p. 169-177.
28. Lange, O.F. and H. Grubmuller, *Generalized correlation for biomolecular dynamics.* Proteins: Structure Function and Bioinformatics, 2006. **62**(4): p. 1053-1061.
29. Brooks, B.R., D. Janezic, and M. Karplus, *Harmonic analysis of large systems. 1. Methodology.* Journal of Computational Chemistry, 1995. **16**(12): p. 1522-1542.
30. Ichiye, T. and M. Karplus, *Collective Motions in Proteins: A Covariance Analysis of Atomic Fluctuations in Molecular Dynamics and Normal Mode Simulations.* Proteins: Structure Function and Genetics, 1991. **11**(3): p. 205-217.
31. Kasson, P.M. and V.S. Pande, *Combining Mutual Information With Structural Analysis To Screen For Functionally Important Residues In Influenza Hemagglutinin.* Pacific Symposium On Biocomputing, 2009: p. 492-503.
32. Kelley, L.A., S.P. Gardner, and M.J. Sutcliffe, *An automated approach for clustering an ensemble of NMR-derived protein structures into conformationally related subfamilies.* Protein Engineering Design and Selection, 1996. **9**(11): p. 1063.
33. Maragakis, P. and M. Karplus, *Large amplitude conformational change in proteins explored with a plastic network model: Adenylate kinase.* Journal of Molecular Biology, 2005. **352**(4): p. 807-822.
34. Brink, J., et al., *Experimental verification of conformational variation of human fatty acid synthase as predicted by normal mode analysis.* Structure, 2004. **12**(2): p. 185-191.
35. Ming, D., et al., *How to describe protein motion without amino acid sequence and atomic coordinates.* Proceedings of the National Academy of Sciences of the United States of America, 2002. **99**(13): p. 8620-8625.
36. Tama, F., W. Wriggers, and C.L. Brooks, *Exploring global distortions of biological macromolecules and assemblies from low-resolution structural information and elastic network theory.* Journal of Molecular Biology, 2002. **321**(2): p. 297-305.
37. Seeger, C. and W.S. Mason, *Hepatitis B virus biology.* Microbiology And Molecular Biology Reviews, 2000. **64**(1): p. 51-68.
38. Caspar, D.L.D. and A. Klug, *PHYSICAL PRINCIPLES IN CONSTRUCTION OF REGULAR VIRUSES.* Cold Spring Harbor Symposia On Quantitative Biology, 1962. **27**: p. 1-24.

39. Gibbons, M.M. and W.S. Klug, *Influence of nonuniform geometry on nanoindentation of viral capsids.* Biophysical Journal, 2008. **95**(8): p. 3640-3649.
40. Lu, M.Y. and J.P. Ma, *The role of shape in determining molecular motions.* Biophysical Journal, 2005. **89**(4): p. 2395-2401.
41. Rawat, U.B.S., et al., *A cryo-electron microscopic study of ribosome-bound termination factor RF2.* Nature, 2003. **421**(6918): p. 87-90.
42. Klaholz, B.P., et al., *Structure of the Escherichia coli ribosomal termination complex with release factor 2.* Nature, 2003. **421**(6918): p. 90-94.
43. de Groot, B.L., G. Vriend, and H.J.C. Berendsen, *Conformational changes in the chaperonin GroEL: New insights into the allosteric mechanism.* Journal Of Molecular Biology, 1999. **286**(4): p. 1241-1249.
44. Yang, Z., P. Majek, and I. Bahar, *Allosteric Transitions of Supramolecular Systems Explored by Network Models: Application to Chaperonin GroEL.* Plos Computational Biology, 2009. **5**(4).
45. Ma, J.P. and M. Karplus, *The allosteric mechanism of the chaperonin GroEL: A dynamic analysis.* Proceedings of the National Academy of Sciences of the United States of America, 1998. **95**(15): p. 8502-8507.
46. Hyeon, C., G.H. Lorimer, and D. Thirumalai, *Dynamics of allosteric transitions in GroEL.* Proceedings of the National Academy of Sciences of the United States of America, 2006. **103**(50): p. 18939-18944.
47. Tehver, R., J. Chen, and D. Thirumalai, *Allostery Wiring Diagrams in the Transitions that Drive the GroEL Reaction Cycle.* Journal Of Molecular Biology, 2009. **387**(2): p. 390-406.
48. Cui, Q. and M. Karplus, *Allostery and cooperativity revisited.* Protein Science, 2008. **17**(8): p. 1295-1307.
49. Sanner, M.F., A.J. Olson, and J.C. Spehner, *Reduced surface: An efficient way to compute molecular surfaces.* Biopolymers, 1996. **38**(3): p. 305-320.
50. Brooks, B.R., et al., *CHARMM—A program for macromolecular energy, minimization, and dynamics calculations.* Journal of Computational Chemistry, 1983. **4**(2): p. 187-217.

**TABLES**

**Table 1. GroEL FE models**

| FE models | PDB-1OEL | EMD-1081 | EMD-1080 | EMD-1095 |
|---|---|---|---|---|
| **Number of nodes** | 50,738 | 40,182 | 72,919 | 72,439 |
| **Number of elements** | 216,950 | 155,039 | 351,941 | 356,521 |
| **Resolution (Å)** | 2.8 | 6 | 11.5 | 25 |
| **Volume (Å$^3$)** | 902,388 | 937,623 | 942,211 | 989,727 |

**FIGURES**

Figure 1. Automated procedure for computation of conformational dynamics using the FE framework.

Figure 2. Discretized molecular surface of kinesin dimers bound to a microtubule (EMD-1030) (A) generated by Chimera at the suggested contour level of 67.1 prior to surface mesh repair and (B) after surface mesh repair using Meshlab. (Removed isolated fragments are highlighted in red.)

Figure 3. EM-NMDB statistics for all EMDB structures (681 entries).

Figure 4. Classification of the EM-NMDB according to biological function for solved structures (452 entries)

Figure 5. Representative EM-NMDB structures with molecular surfaces, lowest normal mode with relative magnitude of the normal mode and corresponding relative strain energy, and the root mean squared fluctuations (red denotes high values while blue denotes low values). (EMD-1231) connector of bacteriophage T7; (EMD-1402) hepatitis B viral capsid; (EMD-1080) GroEL; (EMD-1554) 70S *E. coli* ribosome in the pretranslocation state; and (EMD-1030) kinesin dimers bound to a microtubule.

Figure 6. FE models of the hepatitis B viral capsid. Molecular surfaces are obtained (A) from the atomic structure assuming a Gaussian electron density distribution for each atom and (B) from the cryo-EM map (EMD-1402). Model (A) is obtained from M. M. Gibbons and W. S. Klug (Mechanical and Aerospace Engineering, UCLA).

Figure 7. The three lowest normal modes of the hepatitis B viral capsid with relative magnitude of the normal modes displayed (first, second, and third columns), red denotes large relative displacement while blue denotes small relative displacement) and mean modal overlap between the two FE models (last row). Six neighboring modes are used to compute the mean overlap for each mode.

Figure 8. Relative total RMSF of the capsid of hepatitis B (EMD-1402) with (A) hexamer-centered view and (B) pentamer-centered view. Red denotes large relative RMSF while blue denotes low relative RMSF.

Figure 9. Ribosome-bound termination factor RF2 (EMD-1010). (A) FE mesh and (B) relative RMSF using 20 normal modes (red denotes high and blue denotes low relative RMSF).

Figure 10. (A) Clusters of correlated molecular motions found for the ribosome-bound termination factor RF2 (EMD-1010) using the KGS criterion (7 clusters). (B) Maximally correlated distant clusters from (A) colored in red with the generalized correlation coefficient of 0.6810. The generalized correlation coefficients for other distant clusters are 0.45, 0.58, 0.66, 0.62, 0.57, 0.62, 0.61, 0.57, and 0.58 corresponding to clusters (1, 4), (1, 5), (1, 7), (2, 4), (2, 7), (3, 4), (3, 7), (4, 6), and (4, 7), respectively.

Figure 11. FE models of GroEL obtained from (A) the atomic crystal structure PDB ID 1OEL and three EM resolutions: (B) EMD-1081 at 6 Å resolution; (C) EMD-1080 at 11.5 Å resolution; and (D) EMD-1095 at 25 Å resolution. The 6 Å EM-based model has weak connectivity between the rings of GroEL that is not present in the reference atomic model (Figure S6).

Figure 12. The lowest three normal modes of GroEL with relative magnitude of the normal mode displayed (red denotes large relative displacement while blue denotes small relative displacement) and mean modal overlap between the EMDB-based FE models (EMD-1081, EMD-1080 and EMD-1095) and the PDB-based FE model (PDB-1OEL) of GroEL structures. Six neighboring modes are used to compute the mean overlap for each mode.

Figure 13. Correlations between molecular domains in GroEL (EMD-1080, 28 clusters in total). Each subunit is composed of two clusters (one with the equatorial domain and the other with the apical and the intermediate domains as shown in the top-right figure). Cluster correlation is defined as the mean correlation between the residues in the clusters and cluster separation is defined as $D_{ij}/(R_{g,i} + R_{g,j})$ where $D_{ij}$ denotes the distance between the mean position of cluster $i$ and $j$, and $R_{g,i}$ represents the radius of gyration of cluster $i$.

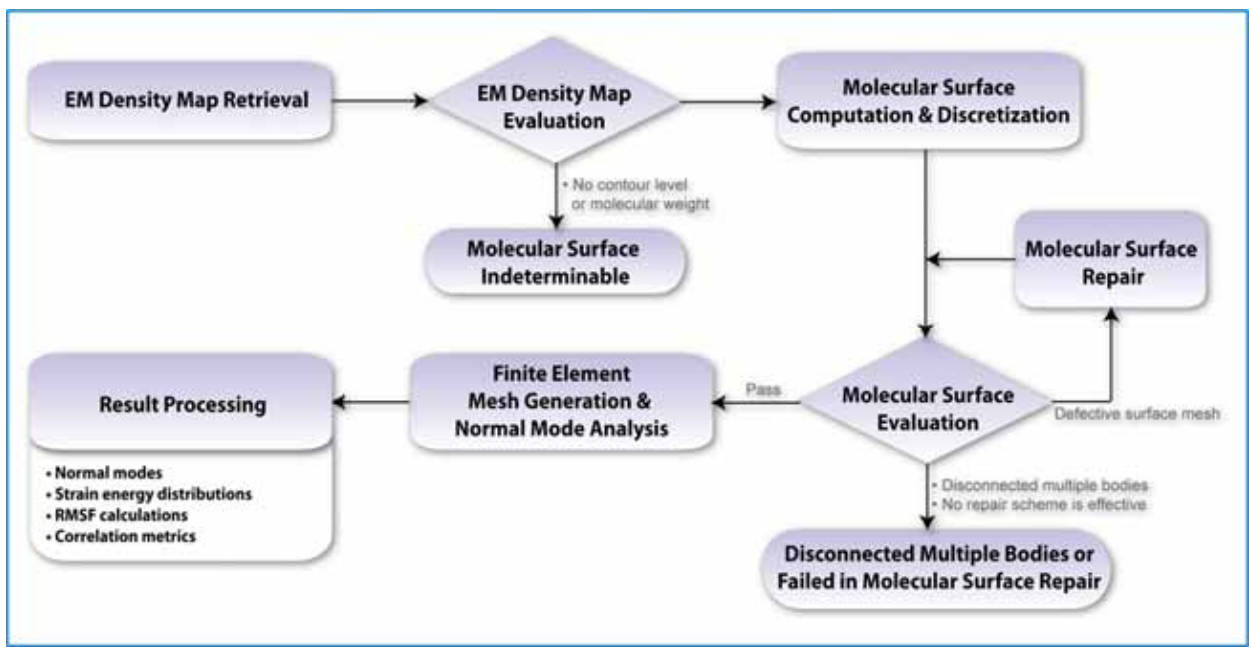

**Figure 1. Automated procedure for computation of conformational dynamics using the FE framework.**

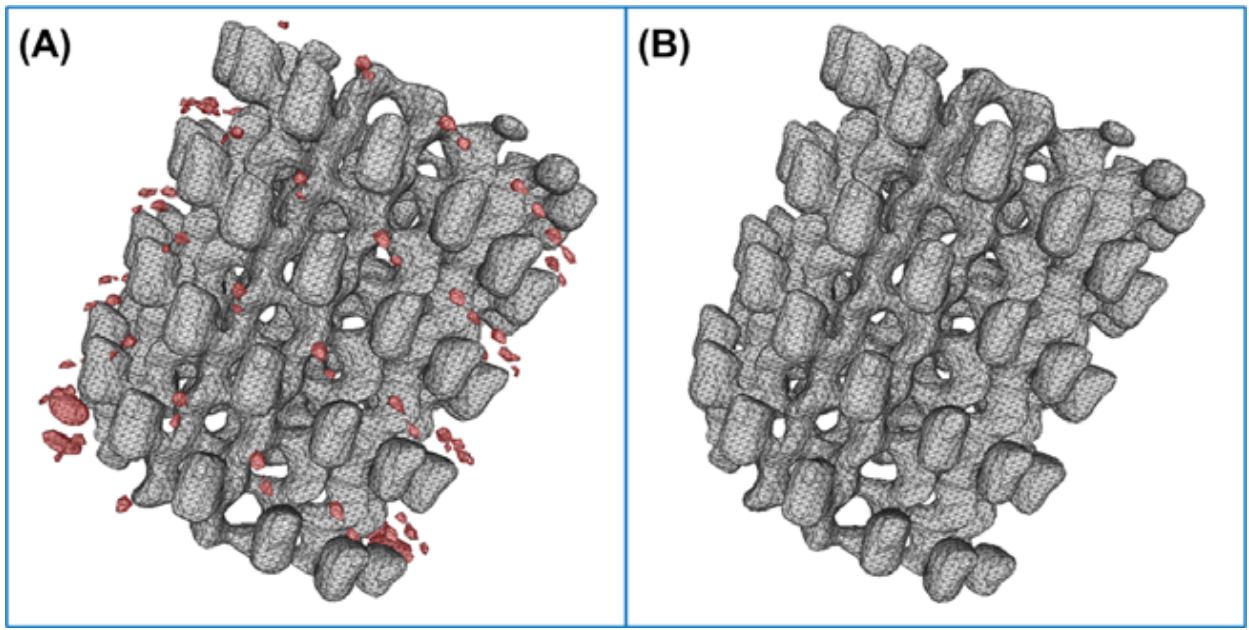

**Figure 2. Discretized molecular surface of kinesin dimers bound to a microtubule (EMD-1030) (A) generated by Chimera at the suggested contour level of 67.1 prior to surface mesh repair and (B) after surface mesh repair using Meshlab. (Removed isolated fragments are highlighted in red.)**

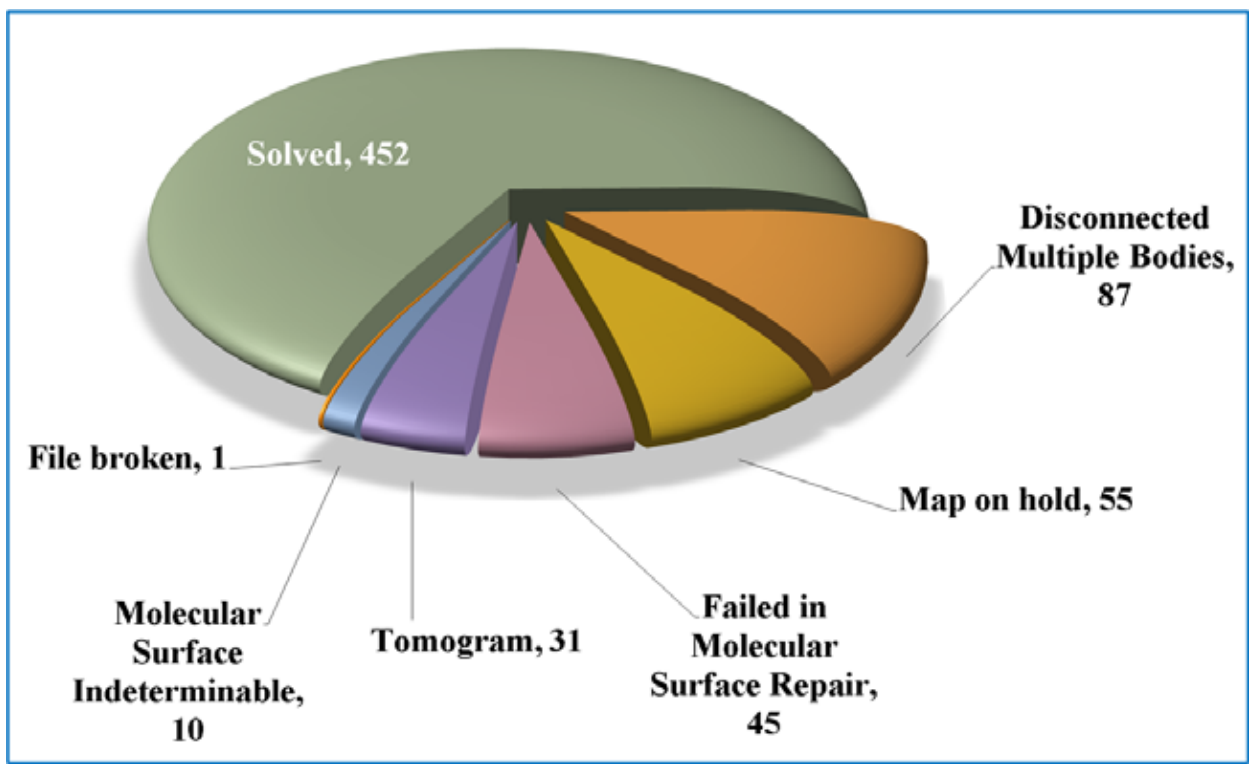

Figure 3. EM-NMDB statistics for all EMDB structures (681 entries).

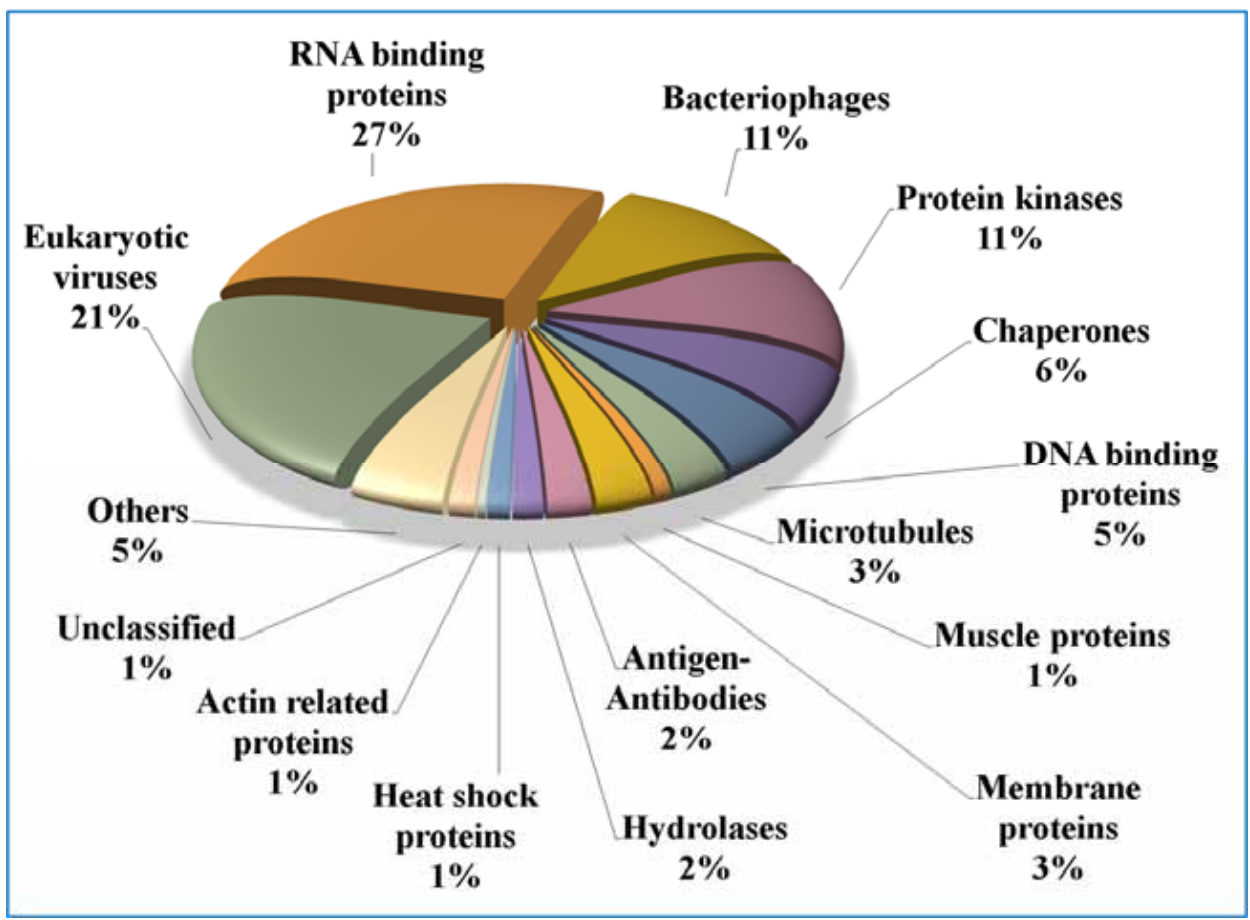

**Figure 4. Classification of the EM-NMDB according to biological function for solved structures (452 entries)**

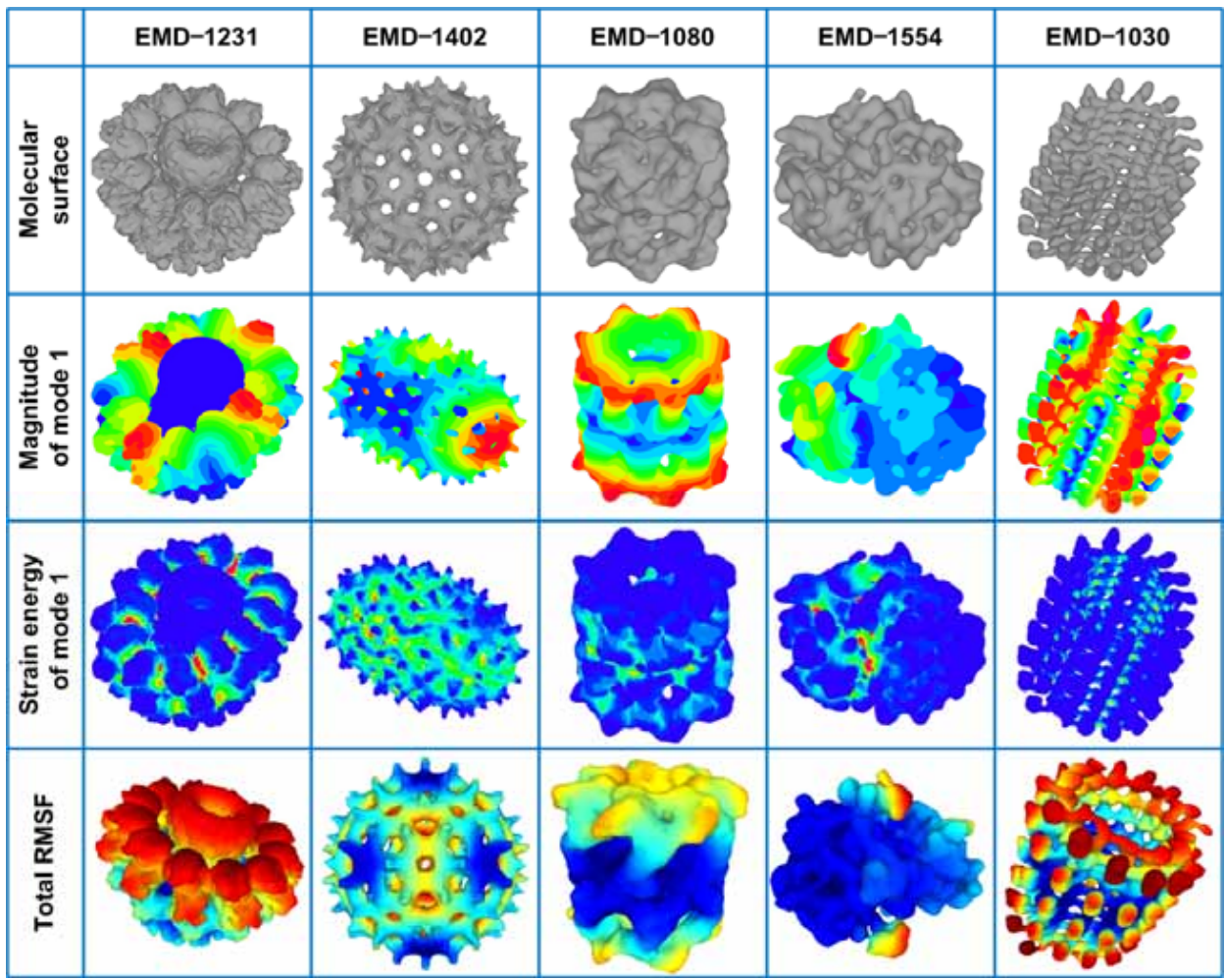

**Figure 5.** Representative EM-NMDB structures with molecular surfaces, lowest normal mode with relative magnitude of the normal mode and corresponding relative strain energy, and the root mean squared fluctuations (red denotes high values while blue denotes low values). (EMD-1231) connector of bacteriophage T7; (EMD-1402) hepatitis B viral capsid; (EMD-1080) GroEL; (EMD-1554) 70S *E. coli* ribosome in the pretranslocation state; and (EMD-1030) kinesin dimers bound to a microtubule.

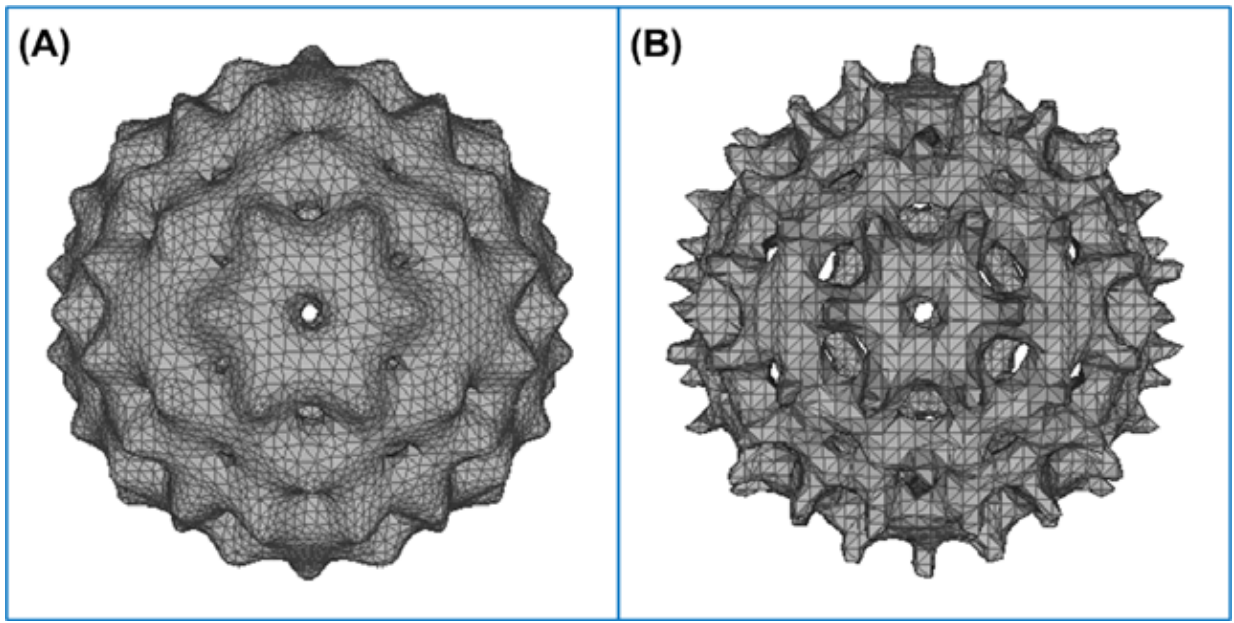

**Figure 6. FE models of the hepatitis B viral capsid. Molecular surfaces are obtained (A) from the atomic structure assuming a Gaussian electron density distribution for each atom and (B) from the cryo-EM map (EMD-1402). Model (A) is obtained from M. M. Gibbons and W. S. Klug (Mechanical and Aerospace Engineering, UCLA).**

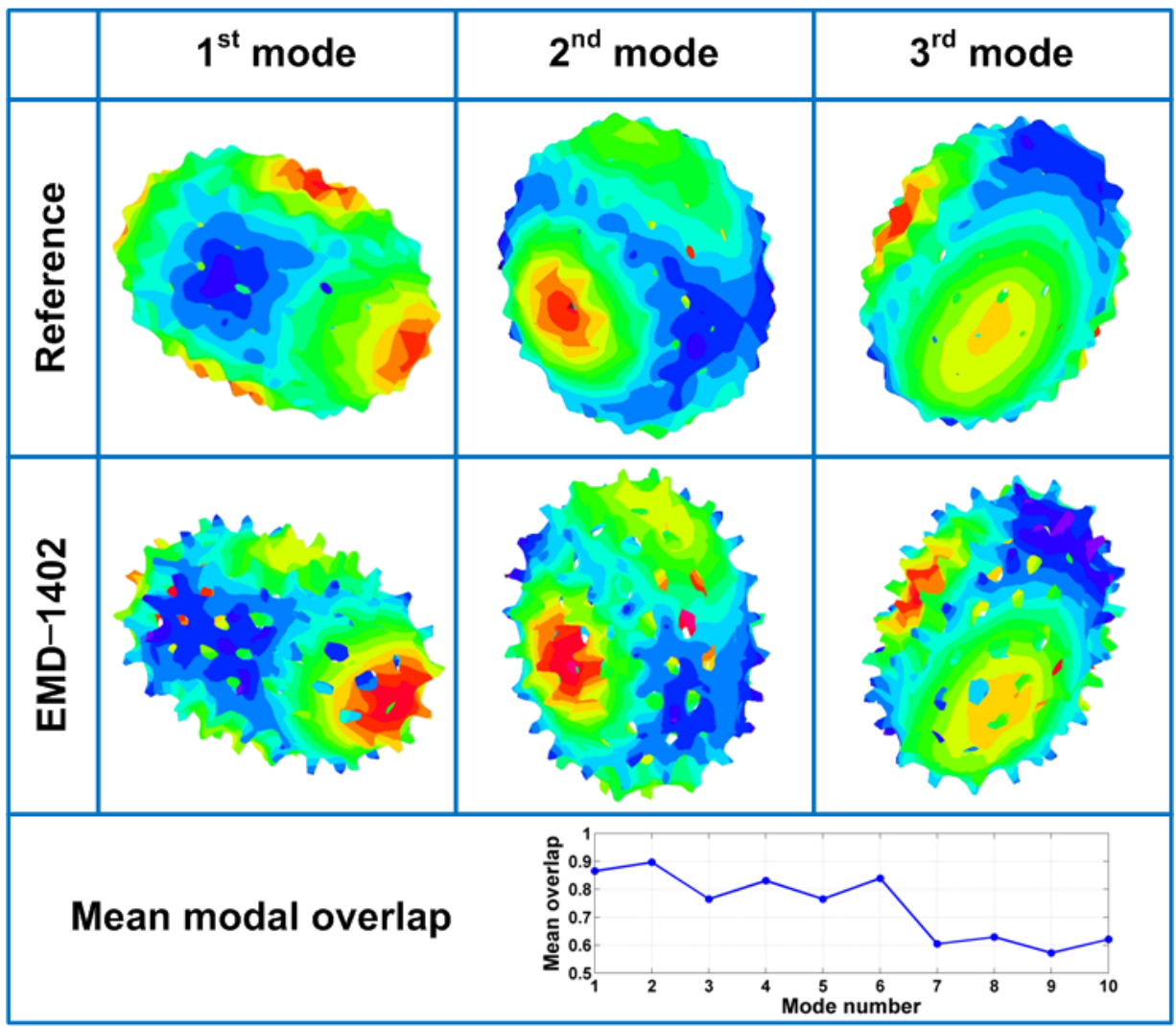

**Figure 7.** The three lowest normal modes of the hepatitis B viral capsid with relative magnitude of the normal modes displayed (first, second, and third columns), red denotes large relative displacement while blue denotes small relative displacement) and mean modal overlap between the two FE models (last row). Six neighboring modes are used to compute the mean overlap for each mode.

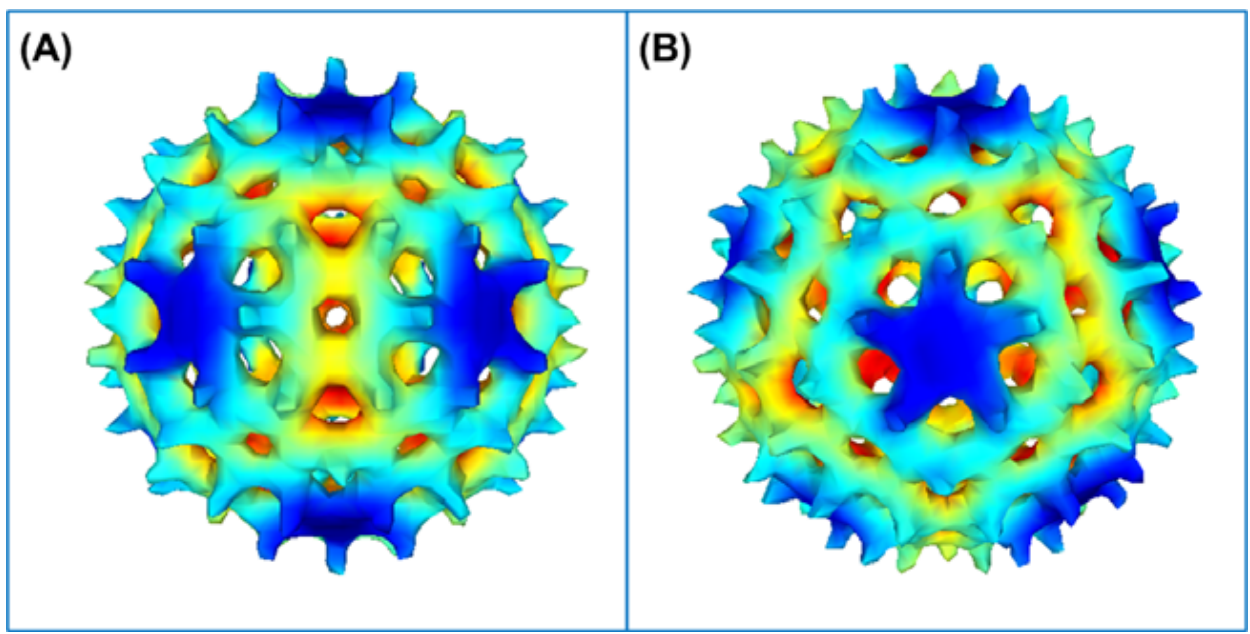

**Figure 8.** Relative total RMSF of the capsid of hepatitis B (EMD-1402) with (A) hexamer-centered view and (B) pentamer-centered view. Red denotes large relative RMSF while blue denotes low relative RMSF.

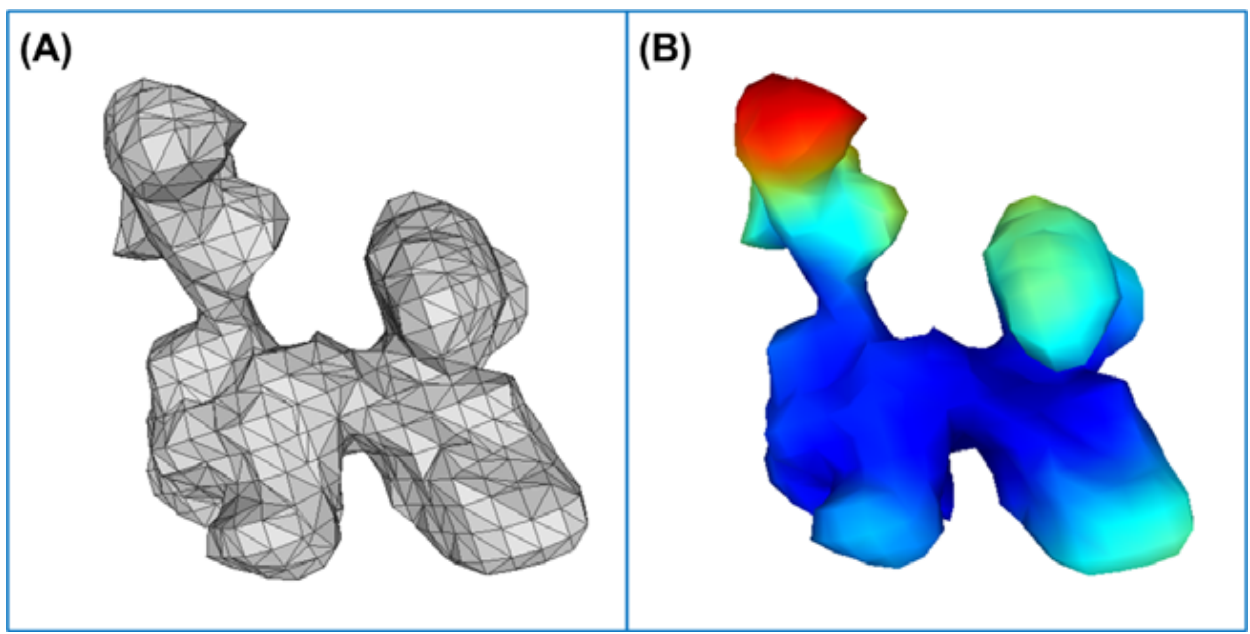

**Figure 9. Ribosome-bound termination factor RF2 (EMD-1010). (A) FE mesh and (B) relative RMSF using 20 normal modes (red denotes high and blue denotes low relative RMSF).**

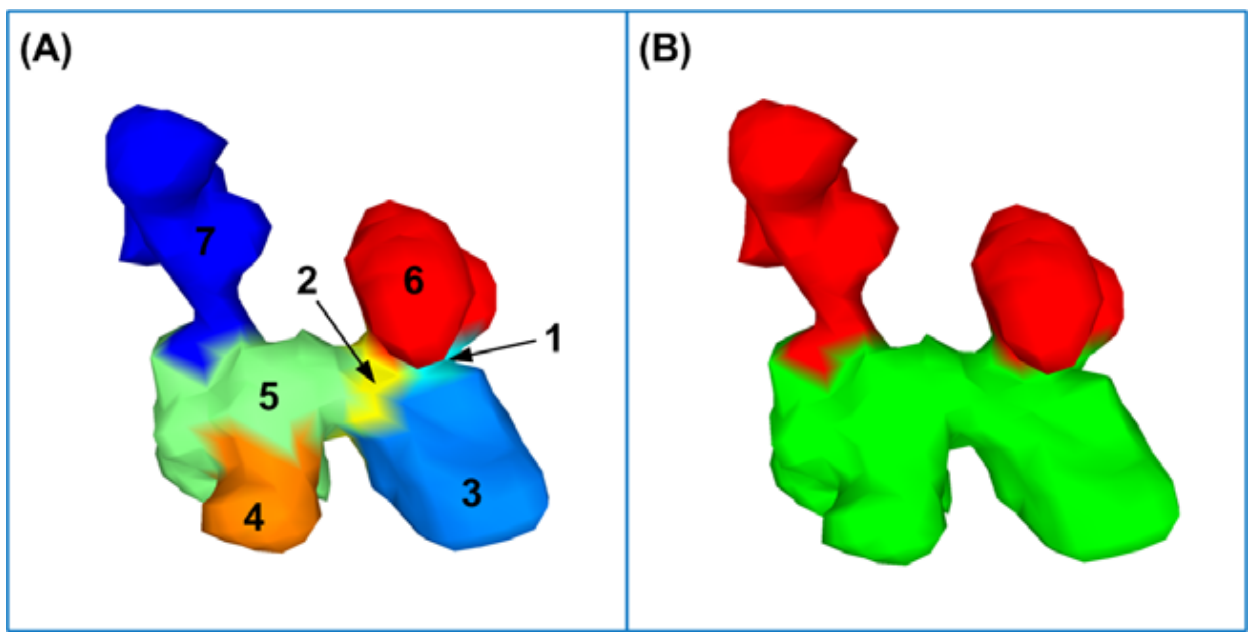

**Figure 10.** (A) Clusters of correlated molecular motions found for the ribosome-bound termination factor RF2 (EMD-1010) using the KGS criterion (7 clusters). (B) Maximally correlated distant clusters from (A) colored in red with the generalized correlation coefficient of 0.6810. The generalized correlation coefficients for other distant clusters are 0.45, 0.58, 0.66, 0.62, 0.57, 0.62, 0.61, 0.57, and 0.58 corresponding to clusters (1, 4), (1, 5), (1, 7), (2, 4), (2, 7), (3, 4), (3, 7), (4, 6), and (4, 7), respectively.

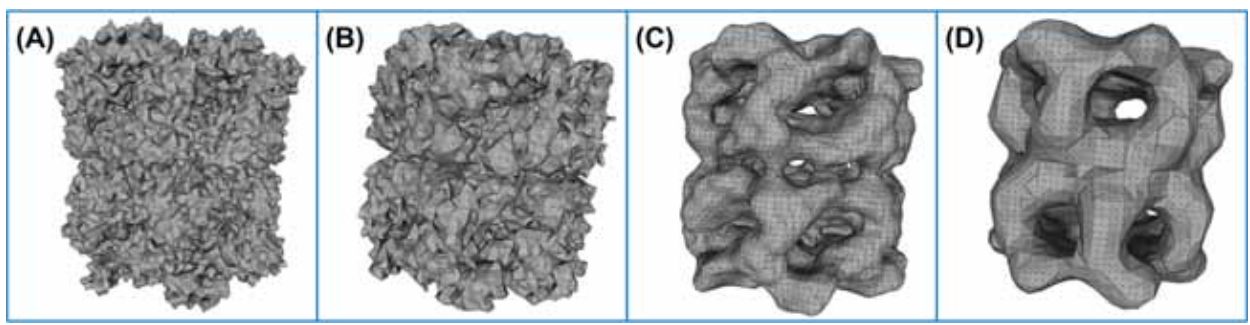

**Figure 11.** FE models of GroEL obtained from (A) the atomic crystal structure PDB ID 1OEL and three EM resolutions: (B) EMD-1081 at 6 Å resolution; (C) EMD-1080 at 11.5 Å resolution; and (D) EMD-1095 at 25 Å resolution. The 6 Å EM-based model has weak connectivity between the rings of GroEL that is not present in the reference atomic model (Figure S6).

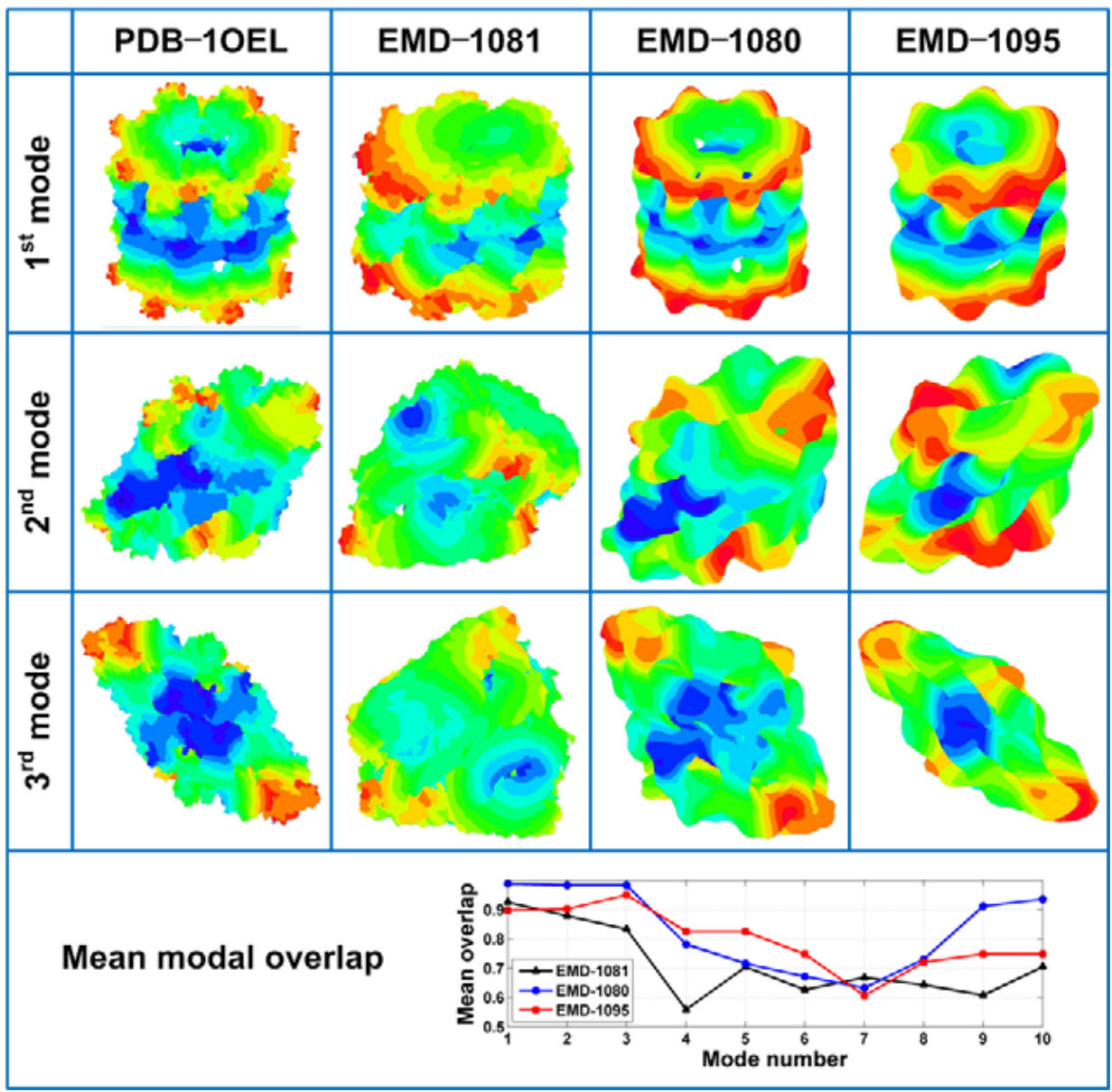

**Figure 12.** The lowest three normal modes of GroEL with relative magnitude of the normal mode displayed (red denotes large relative displacement while blue denotes small relative displacement) and mean modal overlap between the EMDB-based FE models (EMD-1081, EMD-1080 and EMD-1095) and the PDB-based FE model (PDB-1OEL) of GroEL structures. Six neighboring modes are used to compute the mean overlap for each mode.

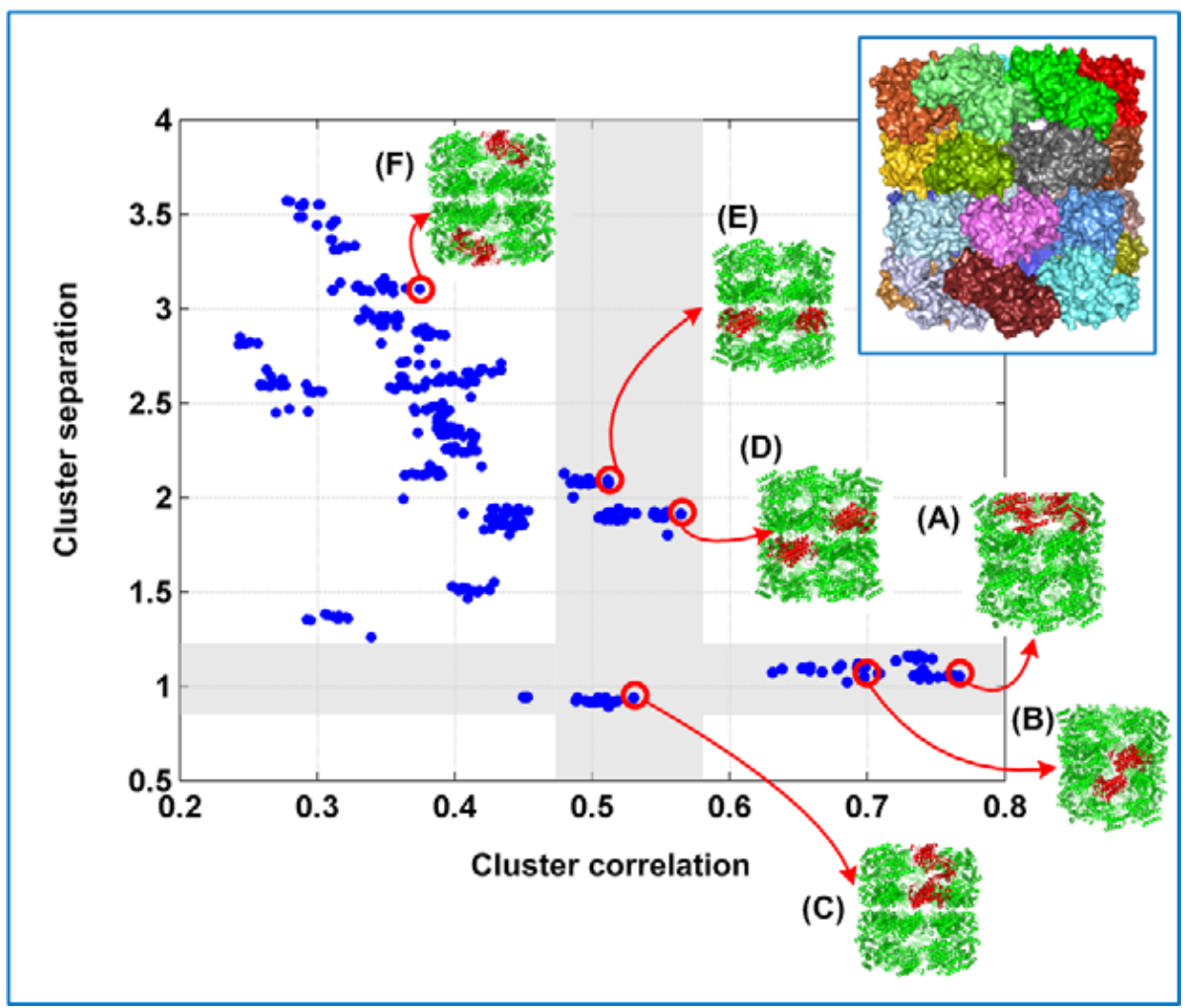

**Figure 13.** Correlations between molecular domains in GroEL (EMD-1080, 28 clusters in total). Each subunit is composed of two clusters (one with the equatorial domain and the other with the apical and the intermediate domains as shown in the top-right figure). Cluster correlation is defined as the mean correlation between the residues in the clusters and cluster separation is defined as $D_{ij}/(R_{g,i}+R_{g,j})$ where $D_{ij}$ denotes the distance between the mean position of cluster $i$ and $j$, and $R_{g,i}$ represents the radius of gyration of cluster $i$.

SUPPLEMENTARY MATERIAL

Table S1. EMDB entries excluded from NMA.

Figure S1. Examples of EMDB entries classified as disconnected multiple bodies and excluded from NMA (colors highlight all or parts of disconnected fragments). (A) EMD-1428 (Microtubule) (B) EMD-5104 (ATPases)  (C) EMD-1111 (Virus)  (D) EMD-1458 (GroEL)  (E) EMD-1073 (Ribosome)  (F)  EMD-1088 (Acrosomal actin bundle)  (G) EMD-1061 (Inositol 1,4,5-triphosphate receptor)  (H) EMD-1335 (Tail of bacteriophage K1-5)  (I) EMD-1591 (Anaphase promoting complex)

Figure S2. Sample EM-NMDB structures. Mobile regions of the lowest normal mode are highlighted in red corresponding to top 20% of mode magnitude. Figures are created using Chimera by importing the original density map (transparent) and the lowest mode magnitude map (red) together. (A) kinesin dimers bound to a microtubule (EMD-1030); (B) a GroEL (EMD-1080); (C) bacteriophage P22 tail machine (EMD-1119); (D) connector of bacteriophage T7 (EMD-1231); (E) human RNA polymerase II (EMD-1283); (F) nitrilase from Rhodococcus rhodochrous J1 (EMD-1313); (G) a chaperonin, cpn60 (EMD-1397); (H) parvovirus capsid (EMD-5105)

Figure S3. The root mean square fluctuations of T4 lysozyme (PDB ID 3LZM) computed using 20, 50, 100 and 200 modes. The inserted graph (top-right) shows the variation of the mean relative RMSF as a function of the number of modes used for the RMSF calculations. The RMSF with 200 modes is chosen as the reference. With 20 modes, 77% of the reference RMSF is obtained on average. Normal modes are obtained from all-atom normal mode analysis implemented in CHARMM [50].

Figure S4. Clusters of correlated residues of T4 lysozyme (PDB ID 3LZM).  (A) All clusters resulted from KGS criterion (14 clusters shown in different colors) and (B) the maximally correlated distant clusters, residues 33-53 and residues 81-90 colored in red, corresponding to residues correlated due to hinge-bending.

Figure S5. Clusters of correlated residues of Adenylate kinase (PDB ID 4AKE).  (A) All clusters resulted from KGS criterion (8 clusters shown in different colors) and (B) the maximally correlated distant clusters, residues 30-73 and residues 113-175 colored in red, corresponding to residues active in the conformational change from its open (PDB ID 4AKE) to its closed states (PDB ID 1AKE).

Figure S6. Cross-sections of where two rings of GroEL are connected. The FE models are obtained from (A) the atomic crystal structure PDB ID 1OEL and three EM resolutions: (B) EMD-1081 at 6 Å resolution; (C) EMD-1080 at 11.5 Å resolution; and (D) EMD-1095 at 25 Å resolution. 6 Å EM-based model (B) presents weaker inter-ring connection compared to other models.

**Table S1. EMDB entries excluded from NMA.**

|  | Number of entries | EMDB ID |
|---|---|---|
| **Molecular surface indeterminable** [1] | 10 | 1087, 1151, 1233, 1259, 1533, 1596, 1599, 1601, 1609, 5037 |
| **Disconnected multiple bodies** [2] | 87 | 1015, 1018, 1021, 1025, 1036, 1042, 1052, 1061, 1073, 1079, 1088, 1101, 1106, 1111, 1112, 1118, 1123, 1134, 1137, 1145, 1165, 1176, 1177, 1203, 1207, 1221, 1226, 1229, 1234, 1236, 1237, 1238, 1244, 1254, 1256, 1267, 1268, 1299, 1314, 1320, 1331, 1333, 1334, 1335, 1340, 1341, 1343, 1353, 1374, 1375, 1377, 1379, 1383, 1385, 1387, 1389, 1401, 1415, 1425, 1427, 1428, 1431, 1437, 1442, 1443, 1447, 1458, 1462, 1469, 1471, 1529, 1531, 1532, 1557, 1579, 1580, 1582, 1591, 1617, 5003, 5010, 5012, 5021, 5022, 5023, 5100, 5104 |
| **Failed in molecular surface repair** [3] | 45 | 1016, 1026, 1060, 1075, 1083, 1113, 1115, 1130, 1133, 1152, 1164, 1179, 1181, 1201, 1206, 1235, 1239, 1264, 1265, 1285, 1309, 1316, 1321, 1354, 1371, 1381, 1392, 1412, 1420, 1441, 1444, 1461, 1480, 1489, 1490, 1503, 1509, 1511, 1544, 1549, 1552, 1581, 1593, 5001, 5038 |

[1] Molecular surface indeterminable: Neither the contour level nor the molecular weight is provided in the EMDB which is necessary for the molecular surface determination.
[2] Disconnected multiple bodies: Structures consist of disconnected multiple bodies obtained using the suggested contour level (Figure S1).
[3] Failed in molecular surface repair: The automated, unsupervised procedure fails in repairing the molecular surfaces. The molecular surfaces may be repaired by applying filters manually or in a supervised way, but it is not tried in this study.

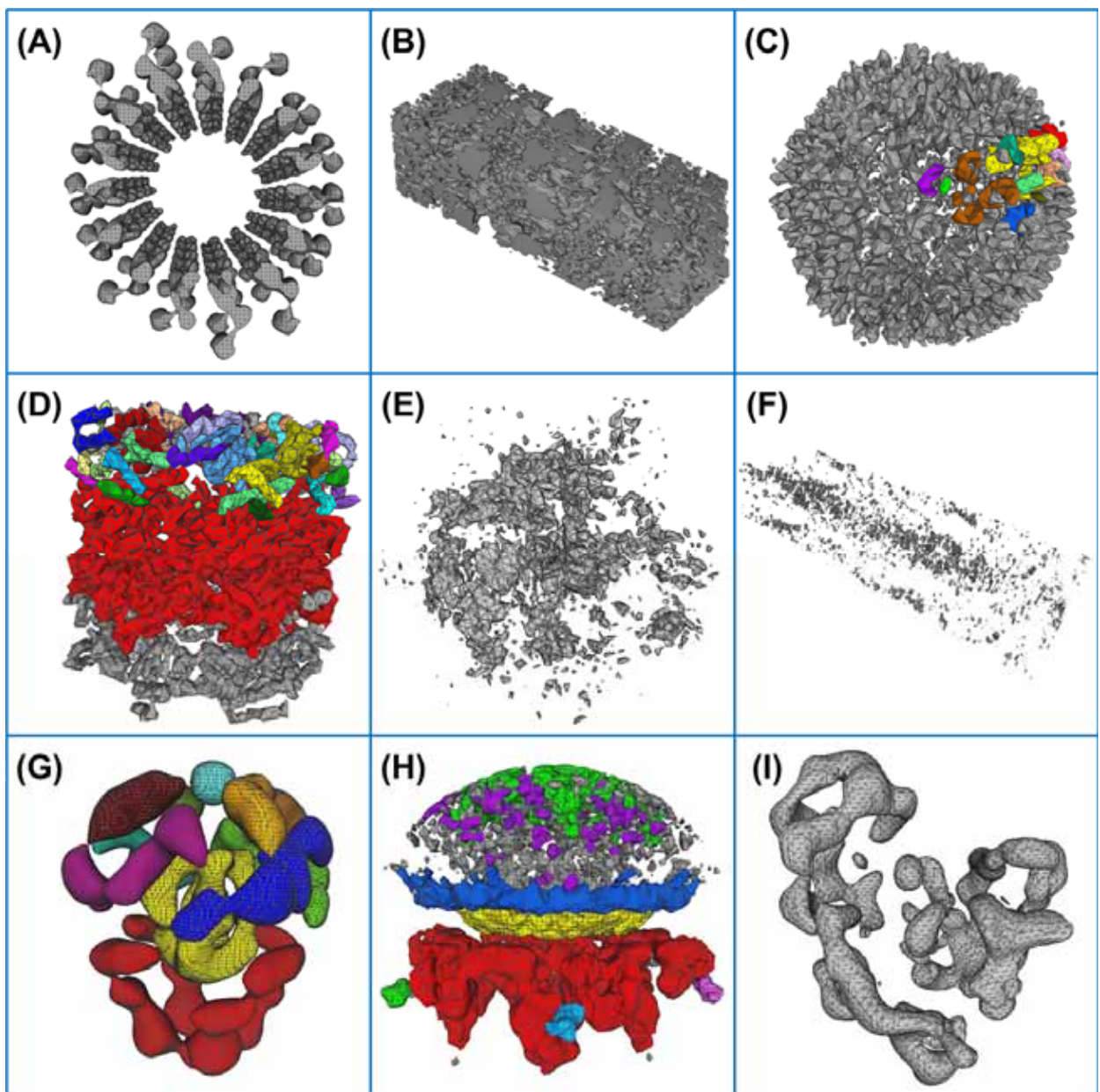

**Figure S1. Examples of EMDB entries classified as disconnected multiple bodies and excluded from NMA (colors highlight all or parts of disconnected fragments). (A) EMD-1428 (Microtubule) (B) EMD-5104 (ATPases) (C) EMD-1111 (Virus) (D) EMD-1458 (GroEL) (E) EMD-1073 (Ribosome) (F) EMD-1088 (Acrosomal actin bundle) (G) EMD-1061 (Inositol 1,4,5-triphosphate receptor) (H) EMD-1335 (Tail of bacteriophage K1-5) (I) EMD-1591 (Anaphase promoting complex)**

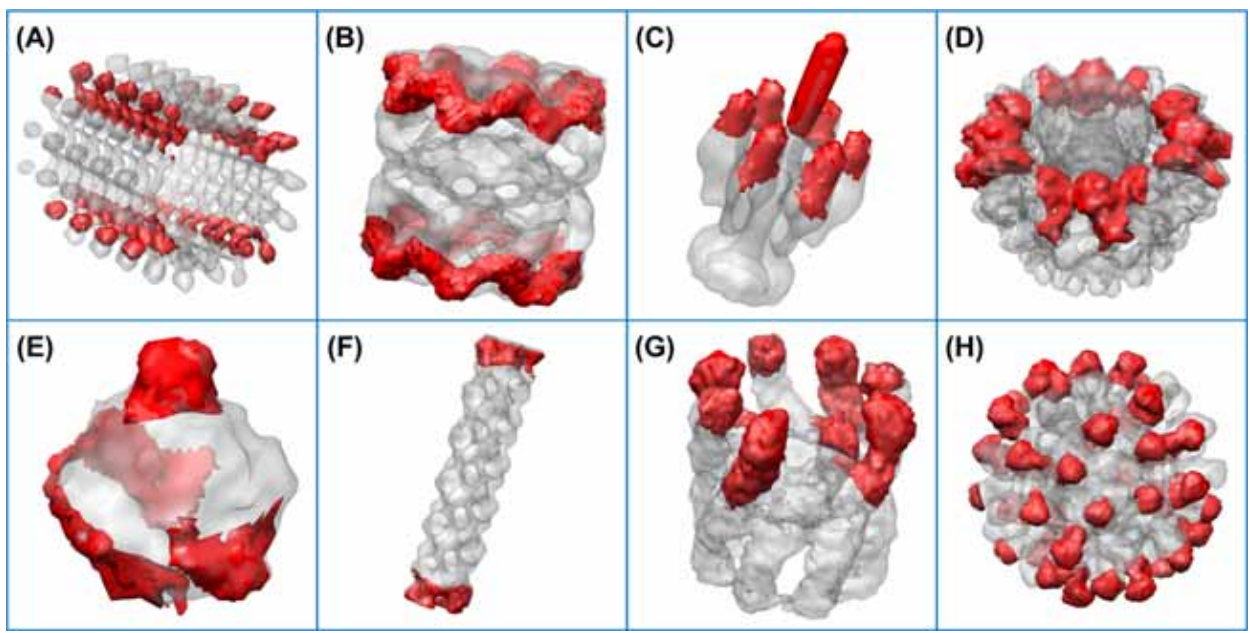

**Figure S2. Sample EM-NMDB structures. Mobile regions of the lowest normal mode are highlighted in red corresponding to top 20% of mode magnitude. Figures are created using Chimera by importing the original density map (transparent) and the lowest mode magnitude map (red) together. (A) kinesin dimers bound to a microtubule (EMD-1030); (B) a GroEL (EMD-1080); (C) bacteriophage P22 tail machine (EMD-1119); (D) connector of bacteriophage T7 (EMD-1231); (E) human RNA polymerase II (EMD-1283); (F) nitrilase from Rhodococcus rhodochrous J1 (EMD-1313); (G) a chaperonin, cpn60 (EMD-1397); (H) parvovirus capsid (EMD-5105)**

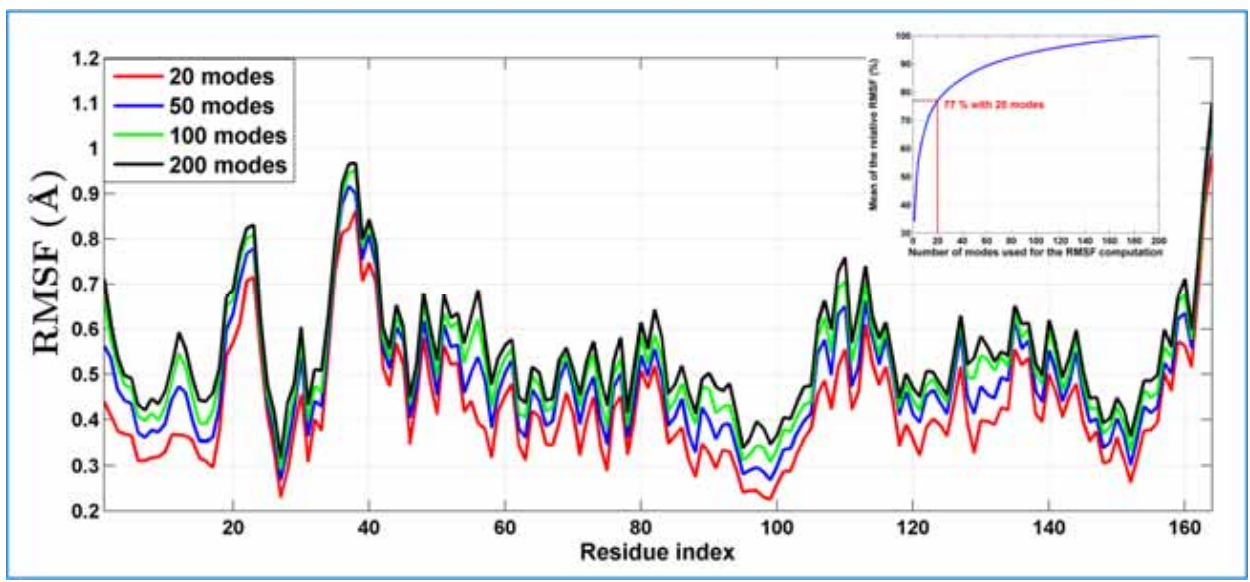

**Figure S3.** The root mean square fluctuations of T4 lysozyme (PDB ID 3LZM) computed using 20, 50, 100 and 200 modes. The inserted graph (top-right) shows the variation of the mean relative RMSF as a function of the number of modes used for the RMSF calculations. The RMSF with 200 modes is chosen as the reference. With 20 modes, 77% of the reference RMSF is obtained on average. Normal modes are obtained from all-atom normal mode analysis implemented in CHARMM [50].

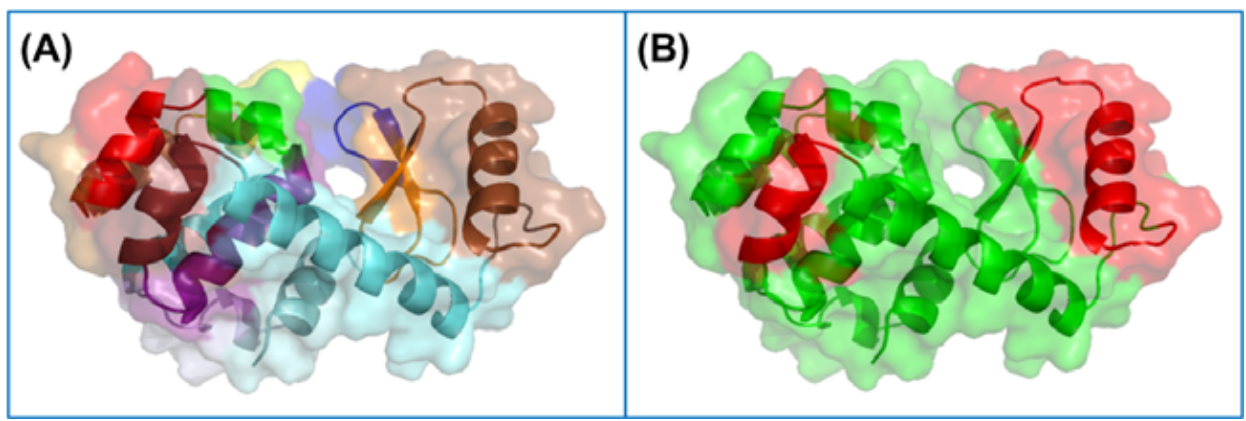

**Figure S4. Clusters of correlated residues of T4 lysozyme (PDB ID 3LZM). (A)** All clusters resulted from KGS criterion (14 clusters shown in different colors) and **(B)** the maximally correlated distant clusters, residues 33-53 and residues 81-90 colored in red, corresponding to residues correlated due to hinge-bending.

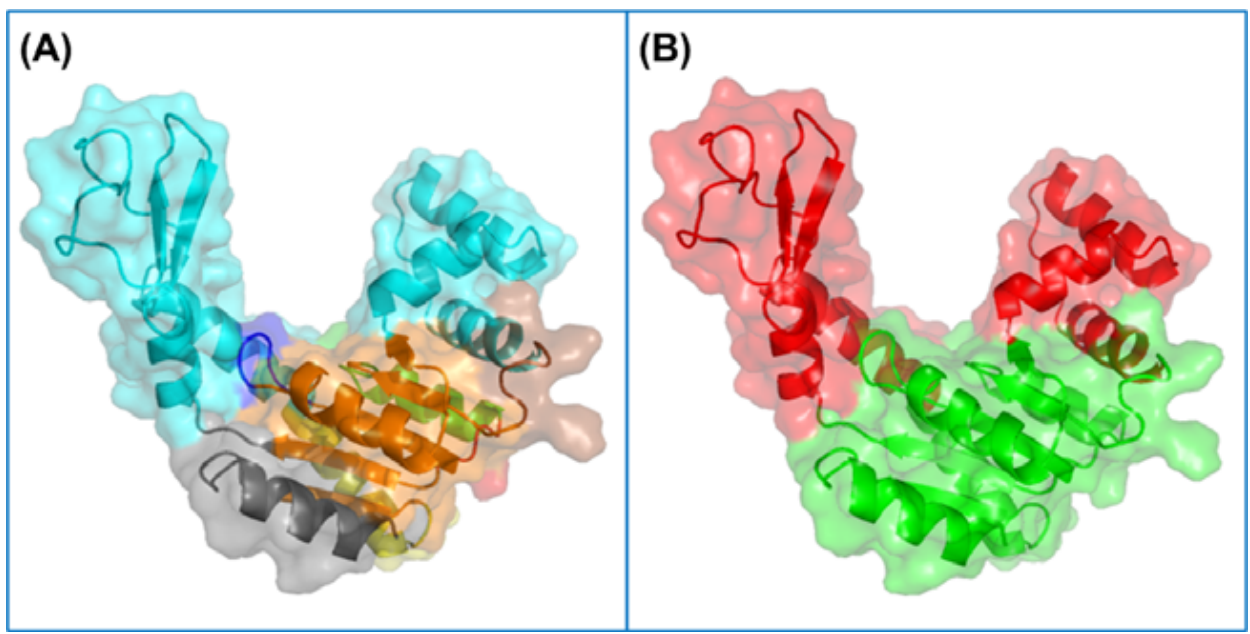

**Figure S5.** Clusters of correlated residues of Adenylate kinase (PDB ID 4AKE). (A) All clusters resulted from KGS criterion (8 clusters shown in different colors) and (B) the maximally correlated distant clusters, residues 30-73 and residues 113-175 colored in red, corresponding to residues active in the conformational change from its open (PDB ID 4AKE) to its closed states (PDB ID 1AKE).

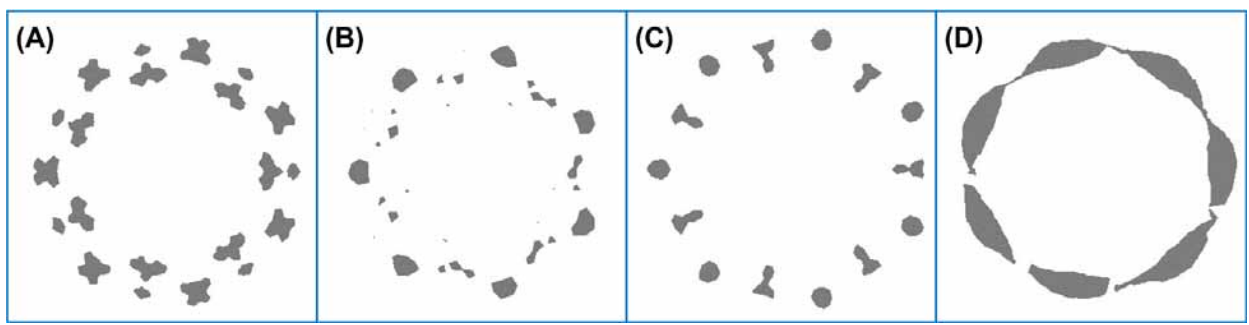

**Figure S6.** Cross-sections of where two rings of GroEL are connected. The FE models are obtained from (A) the atomic crystal structure PDB ID 1OEL and three EM resolutions: (B) EMD-1081 at 6 Å resolution; (C) EMD-1080 at 11.5 Å resolution; and (D) EMD-1095 at 25 Å resolution. 6 Å EM-based model (B) presents weaker inter-ring connection compared to other models.